\documentclass[
superscriptaddress
,preprint
,tightenlines
,showpacs
,showkeys
,nofootinbib
,aps
,amsfonts
,amssymb
,byrevtex
,pdftex %% Allowes to use pdf files as graphics
]{revtex4-1}

\pdfoutput=1
\usepackage{graphicx}
\usepackage{overpic} 
\usepackage{amsmath, amssymb, bm,mathtools} % AMS packages 
\usepackage[dvips]{color}

\usepackage{psfrag,epsfig}
\usepackage{natbib}
\usepackage{hyperref}
\usepackage{slashed}

\usepackage{braket}
\usepackage{float}
\usepackage{subcaption}
\usepackage{ragged2e}
\DeclareCaptionJustification{justified}{\justifying}
\usepackage{multirow}
\usepackage{dcolumn}
\usepackage{booktabs}

\usepackage{csvsimple} %Load CSV file as tables

\usepackage[normalem]{ulem}  %This packet is useful for drafting

\def\){\right)} 
\def\({\left(} 
\def\]{\right]} 
\def\[{\left[}

\def\affKMU{\affiliation{Faculty of Engineering, Karamanoglu Mehmetbey University,\\ Karaman 70100, Turkey}}

\begin{document}

\title{Adiabatic projection method with Euclidean time subspace projection}

\author{Serdar~Elhatisari}
\email{serdarelhatisari@kmu.edu.tr}
\affKMU

\begin{abstract}

Euclidean time projection is a powerful tool that uses exponential decay to extract the low-energy information of quantum systems. The adiabatic projection method, which is based on Euclidean time projection, is a procedure for studying scattering and reactions on the lattice. The method constructs the adiabatic Hamiltonian that gives the low-lying energies and wave functions of two-cluster systems. In this paper we seek the answer to the question whether an adiabatic Hamiltonian constructed in a smaller subspace of the two-cluster state space can still provide information on the low-lying spectrum and the corresponding wave functions. We present the results from our investigations on constructing the adiabatic Hamiltonian using Euclidean time projection and extracting details of the low-energy spectrum and wave functions by diagonalizing it. In our analyses we consider systems of fermion-fermion and fermion-dimer interacting via a zero-range attractive potential in one dimension, and fermion-fermion interacting via an attractive Gaussian potential in three dimensions. The results presented here provide a guide for improving the adiabatic projection method and for reducing the computational costs of large-scale calculations of \emph{ab initio} nuclear scattering and reactions using Monte Carlo methods.

\end{abstract}

\date{\today}
\maketitle

%\tableofcontents

\section{{Introduction}}
\label{sec:intr}

Understanding the structure of atomic nuclei and answering important questions relevant for the synthesis of the elements requires more nuclear data. Experimental efforts are limited due to the fact that important nuclear processes involve charged nuclei and the repulsive Coulomb effects at energies relevant for stellar nucleosynthesis result in inaccurate cross sections~\cite{Weaver:1993zz,Adelberger:2010qa,Leva:2016wmg}. Therefore, \emph{ab initio} calculations of nuclear scattering and reactions are of significant importance for answering long standing questions in nuclear physics. There has been significant progress in \emph{ab initio} methods which combine modern theoretical approaches with powerful computational developments~\cite{Nollett:2006su,Navratil:2011zs,Hagen:2012rq,Quaglioni:2013kma,Elhatisari:2015iga,Navratil:2016ycn,Dohet-Eraly:2015ooa}.  However, in most of these numerical methods the computational scaling limits studying nuclear reactions with a projectile nucleus $A > 3$. Therefore, the computational capabilities of the current \emph{ab initio} methods are needed to be improved to perform large scale calculations.

The Euclidean time projection method is a powerful numerical technique that gives access to the low-energy information of the systems by projecting out the high-energy physics. The fact that makes the Euclidean time projection so powerful is that it works as a subspace projection. See Chapter~6 of Ref.~\cite{lahde2019nuclear} for a detailed discussion. The adiabatic projection method is a standard technique for studying elastic scattering and inelastic reactions of two clusters on the lattice. By means of the Euclidean time projection the adiabatic projection method  describes the low-energy physics of the systems consisting of clusters. The method uses every possible cluster state on the lattice to construct the initial cluster states, $\ket{\bf{R}}$, which is parameterized by the relative separation between clusters, $\bf{R}$, and evolve them in Euclidean time using the microscopic Hamiltonian of underlying theory. The Euclidean time evolution acts as a low-energy filter, and in the limit of large Euclidean projection time, the description of the low-lying two-cluster states becomes exact and they span the low-energy subspace of two-body continuum states for participating clusters. Then these dressed two-cluster states are used to construct a low-energy effective Hamiltonian for clusters, which is called the adiabatic Hamiltonian. Finally, by diagonalizing the adiabatic Hamiltonian we can  obtain the low-energy spectrum and corresponding wave functions of the two-cluster system, and study reactions such as,
\begin{align}
a + b \to a + b, \nonumber \\
a + b \to c + d, \\
a + b \to c + \gamma,  \nonumber 
\end{align}
where $a$, $b$, $c$ are nuclear clusters and $\gamma$ is a photon. The method was used to study the radiative neutron capture reaction $p(n,\gamma)d$~\cite{Rupak:2013aue}, the elastic scattering phase shifts for neutron-deutron $d(n,n)d$~\cite{Pine:2013zja,Elhatisari:2014lka}, and proton-proton fusion process $p(p,e^{+} \nu_{e})d$~\cite{Rupak:2014xza}. To improve the accuracy and efficiency of the method, in Refs.~\cite{Rokash:2015hra,Elhatisari:2016hby} computational advancements in the construction of the adiabatic Hamiltonian and in the extraction of the scattering information from the adiabatic Hamiltonian were introduced. These computational developments helped for the efficiency and accuracy of larger-scale calculations of alpha-alpha scattering using Monte Carlo simulations~\cite{Elhatisari:2015iga,Elhatisari:2016owd}.

The adiabatic Hamiltonian is constructed using the low-energy subspace of two-body continuum states, and its eigenstates are not orthogonal to and contaminated with the states spanning outside this low-energy subspace. Nevertheless, the Euclidean time projection method plays crucial role and releases these contaminations and gives the exact description of the low-lying states for large Euclidean time. This raises the question whether the adiabatic Hamiltonian constructed with a smaller number of initial cluster states defined on the lattice can give the low-energy information. Answering this question is important because the required computational cost for computing the adiabatic Hamiltonian  scales quadratically with the number of initial cluster states. Therefore, in this paper we study two-cluster scattering on the lattice to answer this question, and we present the results from our calculations and analyses.

%%%%%%%%%%%%%%%%%%%%%%%%%%%%%%%%%%%%%%%%%%%%%%%%%%%%%%%%%%%%%%%%%%%
\section{{Lattice formalism}}
\label{sec:latticeformalism}

In this section we introduce the lattice theory of two-component fermions, and we work with natural units where $\hbar = c = 1$. The non-relativistic Hamiltonian in ${\rm D}$-dimensions is, 
%------------ Equation -------------------------
\begin{align}
\hat{H} = &
\sum_{s} \frac{1}{2 m_{s}} \, 
\int  \, d^{{\rm D}}{\bf r} \,
\nabla b^{\dagger}_{s}({\bf r}) \,
\nabla b^{\,}_{s}({\bf r})  \nonumber\\
& +
\frac{1}{2}\sum_{s,s'}  \,
\int\int  \, d^{{\rm D}}{\bf r} \, d^{{\rm D}}{\bf r}\,' \,
b^{\dagger}_{s'}({\bf r}\,') \, b^{\,}_{s'}({\bf r}\,')
\, V_{ss'}({\bf r} - {\bf r}\,')
b^{\dagger}_{s}({\bf r}) \, b^{\,}_{s}({\bf r})
\,,
\end{align}
where $s$ labels the particle species, ${\rm D}$ stands for the number of spatial dimensions, and $b^{\,}_{s}$ ($b^{\dagger}_{s}$) is the annihilation (creation) operator.

In our calculations, we use a lattice that is periodic with the lattice spacing $a$, and we define all physical quantities in lattice units (l.u.) multiplying them by the corresponding powers of $a$. Therefore, the non-relativistic lattice Hamiltonian is, 
%------------ Equation -------------------------
\begin{align}
\hat{H }= \hat{H}_{0} + \hat{V}
\,,
\label{eqn:Hamiltonian-001}
\end{align}
where the free non-relativistic lattice Hamiltonian is,
%------------ Equation -------------------------
\begin{align}
\hat{H}_{0} = & - 
\sum_{s} \frac{1}{2 m_{s}} \, 
\sum_{\hat{l} = \hat{1}}^{\rm \hat{D} }  \,
\sum_{{\bf n}}  \,
\left[
\sum_{k = -N}^{N}
c_{|k|}^{(N)}
b^{\dagger}_{s}({\bf n}) \,
b^{\,}_{s}({\bf n} + k \, \hat{l})
\right]\,,
\label{eqn:Hamiltonian-009}
\end{align}
and the lattice potential is,
%------------ Equation -------------------------
\begin{align}
\hat{V} = 
\frac{1}{2}\sum_{s,s'}  \,
\sum_{{\bf n}\,'}  \, \sum_{{\bf n}}  \,
b^{\dagger}_{s'}({\bf n}\,') \, b^{\,}_{s'}({\bf n}\,')
\, V_{ss'}({\bf n} - {\bf n}\,')
b^{\dagger}_{s}({\bf n}) \, b^{\,}_{s}({\bf n})
\,,
\label{eqn:Hamiltonian-0013}
\end{align}
$n$ labels the lattice sites, and $c_{k}^{(N)}$ are the set of coefficients which gives the Laplace operator on the lattice with the truncation error $\mathcal{O}(a^{2N})$~\cite{Lu:2014xfa,Stellin:2018fkj},
%------------ Equation -------------------------
\begin{align}
c_{k}^{(N)} =
    \begin{cases}
      (-1)^{k+1}\frac{2 \, (N!)^2}{k^2 \, (N+k)! \, (N-k)!}, & k \ne 0 \\
      -2  \sum_{k=1}^{N} \, c_{k}^{(N)}, & k = 0 \,.
    \end{cases}
\label{eqn:Hamiltonian-005}
\end{align}

For convenience we set parameters to values for systems of nuclear physics. We choose the masses $m_{s} = 1$~GeV, and we use values for the interaction strengths that produce considerable amount of scattering in the systems.

\section{{Adiabatic projection method}}
\label{sec:adiabaticprojectionmethod}

\begin{figure}[!ht]
   \centering
   \includegraphics[width=0.4\linewidth]{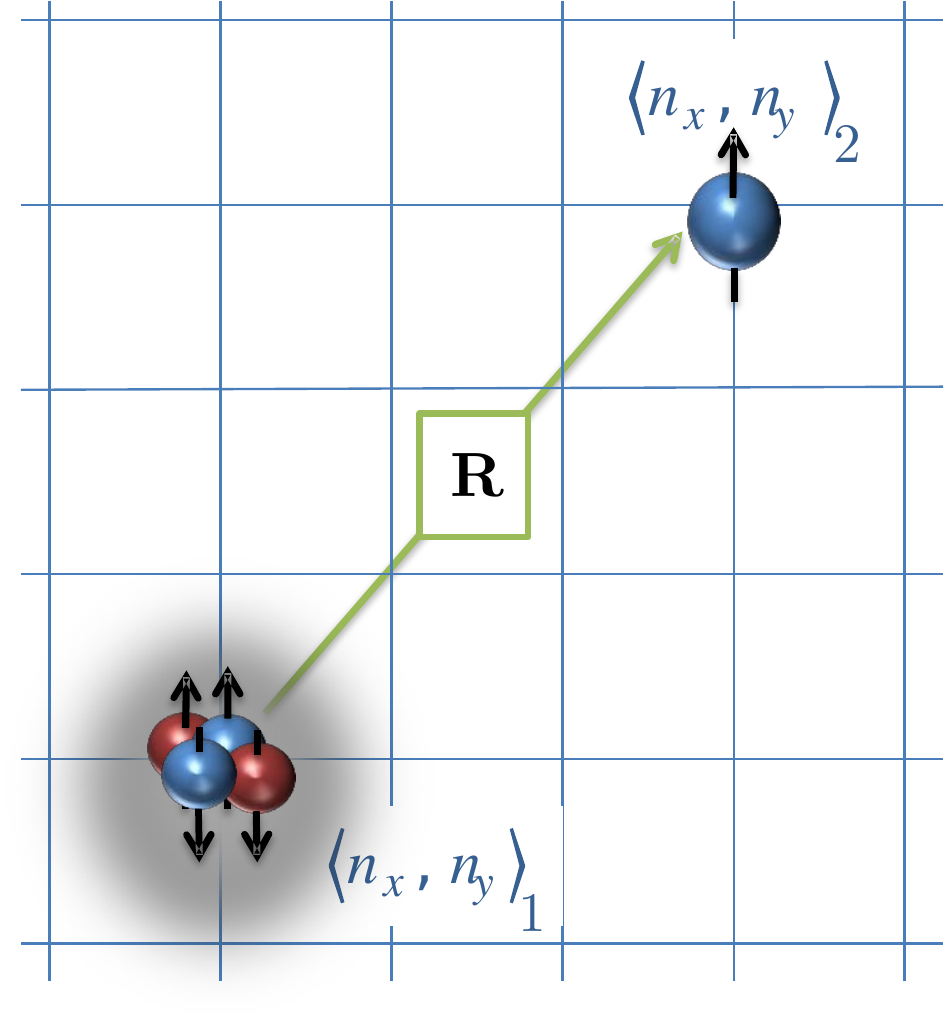}
   \caption{A two-dimensional sketch of the two-cluster initial state $\ket{{\bf R}}$ separated by displacement vector ${\bf R}$}
   \label{fig:initial-cluster-states}
\end{figure}

In this section we discuss the adiabatic projection method which constructs a low-energy effective Hamiltonian for clusters,  the so-called adiabatic Hamiltonian. The method uses the initial cluster states, which is parameterized by the two-cluster displacement vector ${\bf R}$. Using the Slater-determinant we define the initial cluster states as,
%------------ Equation -------------------------
\begin{align}
\ket{\bf R} = \sum_{\bf n} \, \ket{\bf n + R}_{1} \, {\otimes} \,  \ket{\bf n }_{2} \,.
\label{eqn:initial-cluster-states-001}
\end{align}
A schematic view of initial cluster state in a two-dimensional lattice is illustrated in Fig.~\ref{fig:initial-cluster-states}. For a $\rm D$-dimensional lattice, the number of initial cluster states is $N_{R} = L^{\rm D}$.
%------------ Figure -------------------------
Then the method evolves these states in Euclidean time using the microscopic Hamiltonian of the system and forms dressed cluster states that is spanning the low-energy subspace of the two-cluster state space,
%------------ Equation -------------------------
\begin{align}
\ket{\bf R}_{\tau} = \exp(-H \, \tau) \, \ket{\bf R} \,,
\label{eqn:dressed-cluster-states-001}
\end{align}
where $\tau = L_{t} \, a_{t}$ is Euclidean time, $L_t$ is the number of time steps, and $a_{t}$ is the lattice spacing in temporal direction. In the limit of large Euclidean time $\tau$, the dressed cluster states span the low-energy subspace of the two-cluster state space. Now we use these dressed cluster states to compute the projected Hamiltonian,
%------------ Equation -------------------------
\begin{align}
[H_{\tau}]_{{\bf R},{\bf R}^{\prime}} \, = \, 
{}_{\tau/2}{\braket{{\bf R}| H |{\bf R}}}_{\tau/2}
\,,
\label{eqn:HamiltonianMatrix-001}
\end{align}
and since the dressed states are not orthogonal we also compute the norm matrix,
%------------ Equation -------------------------
\begin{align}
[N_{\tau}]_{{\bf R},{\bf R}^{\prime}} \, = \, 
{}_{\tau/2}{\braket{{\bf R}|{\bf R}}}_{\tau/2}
\,.
\label{eqn:NormMatrix-001}
\end{align}
Finally, we define the adiabatic Hamiltonian in the subspace spanned by $N_{R}$ two-cluster states as,
%------------ Equation -------------------------
\begin{align}
[H^{a}_{\tau}]_{{\bf R},{\bf R}^{\prime}} = 
\sum_{{\bf R}^{\prime\prime},{\bf R}^{\prime\prime\prime}} \,
[N^{-1/2}_{\tau}]_{{\bf R},{\bf R}^{\prime\prime}} \,
[H_{\tau}]_{{\bf R}^{\prime\prime},{\bf R}^{\prime\prime\prime}} \,
[N^{-1/2}_{\tau}]_{{\bf R}^{\prime\prime\prime},{\bf R}^{\prime}}
\,,
\label{eqn:adiabaticHamiltonianMatrix-001}
\end{align}
and by diagonalizing $[H^{a}_{\tau}]_{{\bf R},{\bf R}^{\prime}}$ we obtain the  low-energy spectrum and eigenstates of the two-cluster system. Here we denote the eigenstates of the adiabatic Hamiltonian by $\ket{\psi_{i}(\tau)}$ and the corresponding energies are $E^{(a)}_{i}(\tau)$ for $i = 0,1,2,\ldots,N_{R}-1$.

For a $\rm D$-dimensional lattice, the number of eigenstates of the microscopic Hamiltonian $\hat{H}$ is $N = L^{{\rm D}(A-1)}$, where $A$ is the total number of particles in the system. Let $\ket{\phi_{i}}$ be eigenstates of the microscopic Hamiltonian $\hat{H}$ with low-lying energies $E_{i}$ ordered as,
%------------ Equation -------------------------
\begin{align}
E_{0} \le E_{1} \le E_{2} \le ... \,.
\label{eqn:low-lying-spectrum-001}
\end{align}
The eigenstates of the adiabatic Hamiltonian form a subspace of  the complete eigenstates of the microscopic Hamiltonian $\hat{H}$. Therefore, the two-cluster low-lying spectrum can be determined only with an error governed by Euclidean time, 
%------------ Equation -------------------------
\begin{align}
\left| E^{(a)}_{n}(\tau) -  E_{n} \right|  \sim  \mathcal{O}\left(e^{-(E_{N_{R}}-E_{n})\tau}\right)\,,
\label{eqn:low-lying-eigenenergy-009}
\end{align}

\begin{figure}[!ht]
\centering
\includegraphics[width=0.5\linewidth]{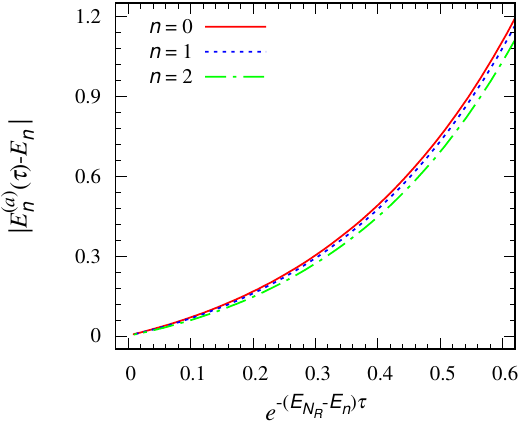}
\caption{Plot shows the error estimates in the lowest three energies of the adiabatic Hamiltonian for a fermion-dimer system. Since $E_{N_{R}} > E_{n}$, the difference $(E_{N_{R}}-E_{n})$ is positive and independent of $\tau$, and for increasing values of $\tau$ the horizontal axes should be read from right to left.}
\label{fig:df_data_errors_2}
\end{figure}

In Eqs.~(\ref{eqn:low-lying-eigenenergy-009}) the difference $(E_{N_{R}}-E_{n})$ is positive and independent of $\tau$, and for increasing Euclidean time $\tau$ the right-hand side of these equations are expected to approach to zero. In Fig.~\ref{fig:df_data_errors_2} for a fermion-dimer system  we show the error estimates in the lowest three energies. The discussion of the fermion-dimer system is the subject of Section~\ref{sec:three-particle-scattering-in-1D}. 
%------------ Figure -------------------------
As seen from Fig.~\ref{fig:df_data_errors_2} as Euclidean time $\tau$ increases the error $e^{-(E_{N_{R}}-E_{n})\tau}$ approaches to zero, and consistently $\left| E^{(a)}_{n}(\tau) -  E_{n} \right|$ vanishes with non-zero slopes. There results indicate that the adiabatic projection in Euclidean time gives a systematically improvable description of the low-lying scattering states of clusters, and in the limit ${\tau \to \infty}$ the description is exact.

In practice, due to the required computational cost it is not practical to go to very large values of $\tau$, and we obtain the exact spectrum  from the adiabatic Hamiltonian by means of Euclidean time extrapolation using an ansatz resembling Eq.(\ref{eqn:low-lying-eigenenergy-009}),
%------------ Equation -------------------------
\begin{align}
E_{n}^{(a)}  = \lim_{\tau \to \infty} E_{n} \,.
\label{eqn:low-lying-eigenenergy-005}
\end{align}

The adiabatic projection method acts as a low-energy filter, and projects out the states corresponding to the high-energy spectrum. Then the question arises, whether using a number of initial cluster states that is less than $N_{R}$ would be sufficient to extract the correct low-energy information from the constructed adiabatic Hamiltonian. If one is interested in extracting only the lowest $N'_{R}$ eigenenergies, where $N'_{R}<N_{R}$, then it might be sufficient that the number of initial cluster states is greater than or equal to $N'_{R}$. Before we proceed to answer this question, for completeness we introduce the radial adiabatic Hamiltonian in order to project onto spherical partial waves.

In three dimensions the initial cluster states can be projected onto spherical harmonics with angular quantum numbers $\ell$, $\ell_{z}$,
%------------ Equation -------------------------
\begin{align}
\ket{R}^{{\rm D} = 3} = \sum_{{\bf R}'} \, Y_{\ell\, \ell_{z}}(\hat{\bf R}') \, \delta_{R^2,{\bf R'}^2} \ket{\bf R'} \,.
\label{eqn:initial-cluster-states-005}
\end{align}
The projection given in Eq.~(\ref{eqn:initial-cluster-states-005}) also ensures that cubic lattice points with the same radial distance squares, $R^{2}$, are summed  by weighting them with the spherical harmonics $Y_{\ell\, \ell_{z}}(\hat{\bf R})$. Furthermore, we can define a radial bin size, $a_{R}$, in the radial direction, then cubic lattice points within the bin size  $a_{R}$ in radial distance can be group together. For a comprehensive discussion and details see Refs.~\cite{Elhatisari:2015iga,Lu:2015riz,Elhatisari:2016hby}. In one-dimensional lattice Eq.~(\ref{eqn:initial-cluster-states-005}) takes rather simple form,
%------------ Equation -------------------------
\begin{align}
\ket{R}^{{\rm D} = 1} = \sum_{{\bf R}'} \, \delta_{R,|{\bf R'}|} \ket{\bf R'} \,.
\label{eqn:initial-cluster-states-011}
\end{align}
and lattice points with equidistant are binned together.
Then the radial dressed cluster states read, 
%------------ Equation -------------------------
\begin{align}
\ket{R}_{\tau}^{{\rm D}} = \exp(-H \, \tau) \, \ket{R}^{{\rm D}} \,,
\label{eqn:dressed-cluster-states-005}
\end{align}
and they are used to compute the projected radial Hamiltonian, norm matrix, and radial adiabatic Hamiltonian in a similar fashion described as Eqs.~(\ref{eqn:HamiltonianMatrix-001}), (\ref{eqn:NormMatrix-001}) and (\ref{eqn:adiabaticHamiltonianMatrix-001}).
%------------ Equation -------------------------
\begin{align}
[H^{a}_{\tau}]_{R,R^{\prime}} = 
\sum_{R^{\prime\prime},R^{\prime\prime\prime}} \,
[N^{-1/2}_{\tau}]_{R,R^{\prime\prime}} \,
[H_{\tau}]_{R^{\prime\prime},R^{\prime\prime\prime}} \,
[N^{-1/2}_{\tau}]_{R^{\prime\prime\prime},R^{\prime}}
\,.
\label{eqn:adiabaticHamiltonianMatrix-005}
\end{align}
After constructing the adiabatic Hamiltonian in radial coordinates, we employ the spherical wall method~\cite{Borasoy:2007vy,Rokash:2015hra,Elhatisari:2016hby}, which is a general method to extract the two-cluster asymptotic wave functions and scattering information efficiently on the lattice. The method introduces a spherical hard wall with radius $R_{\rm wall}$ into the relative separation between two clusters on a periodic lattice and removes the periodic boundary condition. The spherical wall method allows one to efficiently and accurately compute scattering phase shifts or reaction amplitudes for the systems with higher orbital angular momenta, spin-orbital coupling, or in the presence of long-range interactions.

Now we return back to the question whether it is necessary to use all possible initial cluster states parameterized by the relative displacement of two clusters to extract the two-cluster asymptotic wave functions and the energies.  To answer the question one can redefine the initial cluster states given in Eqs.~(\ref{eqn:initial-cluster-states-005}) and (\ref{eqn:initial-cluster-states-011}) by introducing a constraint on the selection of states $\ket{{\bf R}'}$ in the summation. Therefore, we consider the  following initial cluster states in one and three dimensions, respectively,
%------------ Equation -------------------------
\begin{align}
\ket{\mathcal R}^{{\rm D} = 1} = \sum_{{\bf R}'} \, \delta_{R,|{\bf R'}|}  \,\, \delta_{0, \mathrm{mod}(|{\bf R'}|,m)} \ket{\bf R'} \,,
\label{eqn:initial-cluster-states-009}
\end{align}
and
%------------ Equation -------------------------
\begin{align}
\ket{\mathcal R}^{{\rm D} = 3} = \sum_{{\bf R}'} \, Y_{\ell\, \ell_{z}}(\hat{\bf R}') \, \delta_{R^2,{\bf R'}^2}  \,\, \delta_{0, \mathrm{mod}({\bf R'}^2,m)} \ket{\bf R'} \,,
\label{eqn:initial-cluster-states-015}
\end{align}
where $\mathrm{mod}(N,m)$ is the modulo operator. The second Kronecker deltas in Eqs.~(\ref{eqn:initial-cluster-states-009}) and (\ref{eqn:initial-cluster-states-015}) select and add the only every other $m^{\rm th}$ states into the binning. This constraint reduces the number of initial cluster states used in the calculation by a factor of $1/m$ without reducing the range of the two-cluster separation length which is important since we extract the asymptotic wave function from the adiabatic Hamiltonian. When we set $m = 1$, Eq.~(\ref{eqn:initial-cluster-states-009}) and (\ref{eqn:initial-cluster-states-015}) are equivalent to Eqs.~(\ref{eqn:initial-cluster-states-005}) and (\ref{eqn:initial-cluster-states-011}), respectively. Therefore, since we aim to construct the adiabatic Hamiltonians which are restricted to the subspace of dressed cluster states given in Eq.~(\ref{eqn:dressed-cluster-states-001}), in our calculations we use the initial cluster states given in Eqs.~(\ref{eqn:initial-cluster-states-009}) and (\ref{eqn:initial-cluster-states-015}) by setting $m = 2$ or $m = 3$.

\section{Lattice calculations and results}
\label{sec:lattice-calcultions-results}

We study the extraction of the low-lying spectrum and corresponding wave functions from a radial adiabatic Hamiltonian constructed using a set of initial cluster states defined in a subspace of the complete two-cluster separation state space. In the following we consider scattering of two particles in one and three spatial dimensions, and fermion-dimer scattering in one spatial dimension on the periodic lattice. We benchmark the results against scattering parameters from the exact calculations in the continuum limit.

\subsection{Two-particle scattering}
\label{sec:two-particle-scattering}

In this section we consider a simple system of two distinguishable spin--$\frac{1}{2}$ particles with equal masses and interacting via an attractive delta function potential on a periodic lattice. The free non-relativistic microscopic Hamiltonian is defined by Eq.~(\ref{eqn:Hamiltonian-009}), and the delta function potential on the lattice is  
%------------ Equation -------------------------
\begin{align}
\hat{V} = C_{0}
\sum_{{\bf n}}  \,
b^{\dagger}_{\uparrow}({\bf n}) \, b^{\,}_{\uparrow}({\bf n})
b^{\dagger}_{\downarrow}({\bf n}) \, b^{\,}_{\downarrow}({\bf n})
\,,
\label{eqn:Hamiltonian-0015}
\end{align}
where $C_{0}$ is the interaction strength. In this calculation, we set the lattice spacing $a=0.01$~MeV$^{-1}$ and the interaction strength $C_{0}=-0.103$ which forms a dimer with binding energy $2.5$. 
%------------ Figure -------------------------
\begin{figure}[!ht]
\centering
\includegraphics[width=0.7\linewidth]{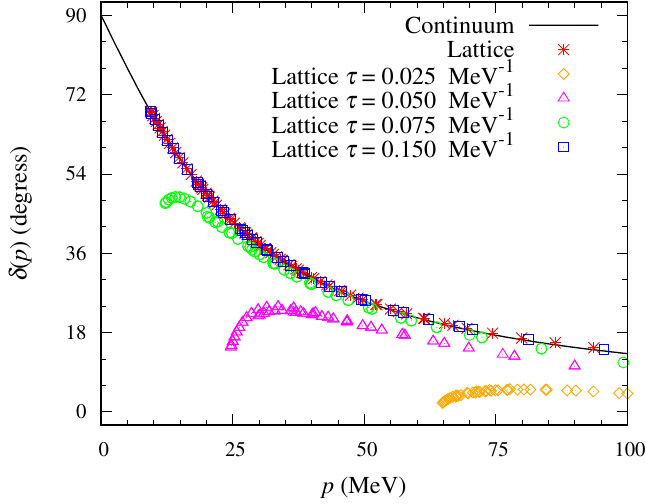}
\caption{Two-particle scattering phase shifts in one spatial dimension calculated three different methods; the exact continuum formula, the microscopic Hamiltonian, and the adiabatic projection method. The solid line is from the continuum calculation. The results from the microscopic Hamiltonian are denoted by red asterisks. The orange diamonds, magenta triangles, green circles and blue squares show the results from the radial adiabatic Hamiltonian at Euclidean time $\tau = 0.025, 0.050, 0.075, 0.150$~MeV$^{-1}$, respectively.}
\label{fig:lattice_vs_BA}
\end{figure}
%------------ Figure -------------------------
\begin{figure}[!h]
\centering
\includegraphics[width=1\linewidth]{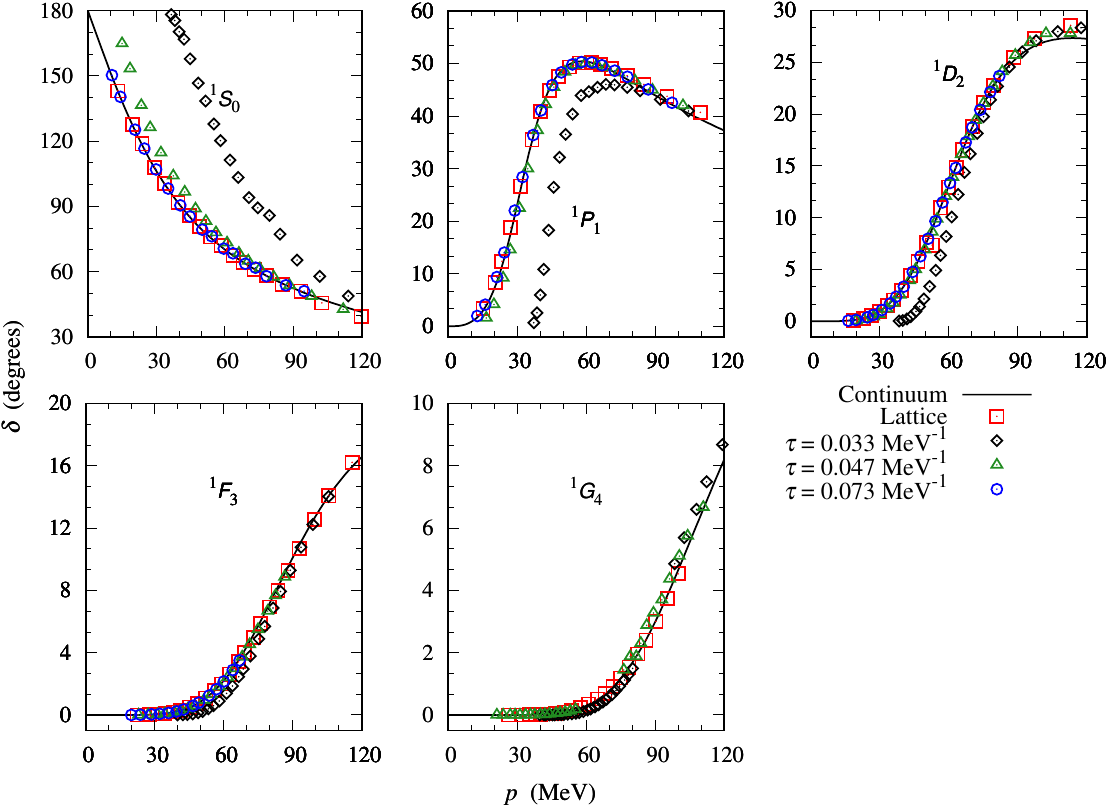}
\caption{Plots of two-particle scattering phase shifts up to $\ell = 4$ in three spatial dimensions. The solid line is the results in the continuum limit. The scattering phase shifts computed using the microscopic Hamiltonian are denoted by red squares. The black diamonds, green triangles, blue circles show the results from the reduced radial adiabatic Hamiltonian at Euclidean time $\tau = 0.033, 0.047, 0.073$~MeV$^{-1}$, respectively.}
\label{fig:two_body_dvsp}
\end{figure}
We construct the radial adiabatic Hamiltonian using initial cluster states given in Eq.~(\ref{eqn:initial-cluster-states-009}) for $m = 2$. The adiabatic projection method, when applied to a system consisting of two point-like particles using all possible two-body initial cluster states given in Eq.~(\ref{eqn:initial-cluster-states-001}), constructs an adiabatic Hamiltonian which is equivalent to the microscopic Hamiltonian. However, here we use $N'_{R}$ initial cluster states, which form a proper subspace of the complete two-cluster separation state space. As a result, the eigenstates of the constructed radial adiabatic Hamiltonian are not completely orthogonal to all two-body states, and they are contaminated with states lying outside the considered subspace. One practical way to reduce these contaminations systematically is to use Euclidean time projection method, which gives access to eigenstates that span the low-energy subspace of the complete system.

We consider a simple system of two particles for simplicity and clarity, and we employ the Euclidean time projection method to construct the radial adiabatic Hamiltonian. Then we compute the scattering phase shifts from the radial adiabatic Hamiltonian at various values of Euclidean time, and these results are benchmarked against phase shifts from the exact calculation in the continuum limit given by,
\begin{align}
\delta(p) = \cot^{-1}\left(-\frac{p}{C_{0} \, \mu}\right),
\label{eqn:phaseshifts-continuum-001}
\end{align}
where $p$ is the relative momentum and $\mu$ is the reduced mass of two-particle system. In Fig.~\ref{fig:lattice_vs_BA} we show two-particle scattering phase shifts in the continuum limit, computed from the microscopic Hamiltonian, and computed from the radial adiabatic Hamiltonian at various values of Euclidean time. These results indicate that the low-lying spectrum of the adiabatic Hamiltonian includes some contributions from the high-energy states of the system, and as the initial states are projected in Euclidean time, it systematically improves the description of the two-particle low-energy states.

To extend our investigation to three spatial dimensions and higher partial wave scattering states, now we focus on a system of two distinguishable spin--$\frac{1}{2}$ particles with equal
masses and interacting via an attractive Gaussian potential on a periodic cubic lattice.
%------------ Equation -------------------------
\begin{align}
\hat{V} = C_{g}
\sum_{{\bf n'},{\bf n}}  \,
b^{\dagger}_{\uparrow}({\bf n'}) \, b^{\,}_{\uparrow}({\bf n'})
\exp\left(-\frac{|\bf{n'}-\bf{n}|^2}{2\,R_{g}^{2}}\right)
b^{\dagger}_{\downarrow}({\bf n}) \, b^{\,}_{\downarrow}({\bf n})
\,,
\label{eqn:Hamiltonian-0019}
\end{align}
where $C_{g}$ is the interaction strength, and $R_{g}$ is the range of the potential $R_{g}$. In this calculation, the lattice spacing is $a = 0.01$~MeV$^{-1}$, the interaction strength is set to $C_{g}=-8.0$~MeV, and the range of the potential is $R_{g} = 0.02$~MeV$^{-1}$. We construct the reduced radial adiabatic Hamiltonian using the initial cluster states defined in Eq.~(\ref{eqn:initial-cluster-states-015}) and $m = 2$ for orbital angular momentum $\ell \le 4$, and we extract the scattering phase shifts using the spherical wall method. The results are summarized in Fig.~\ref{fig:two_body_dvsp}.

\subsection{Fermion-dimer scattering}
\label{sec:three-particle-scattering-in-1D}

Now we extend our investigation beyond two particles and we consider a fermion-dimer system in one spatial dimension consisting of two-component fermions. Here we set the lattice spacing $a=0.01$~MeV$^{-1}$, and we choose the interaction between the two spin components to be attractive and zero-ranged with a strength $C_{0}=-0.103$. This interaction forms a bound state composed of one spin-$\uparrow$ and one spin-$\downarrow$ fermions which is called dimer. Therefore, the fermion-dimer system is the simplest nontrivial example of two-cluster system that one can consider.
%------------ Figure -------------------------
\begin{figure}[!ht]
\centering
\includegraphics[width=0.7\linewidth]{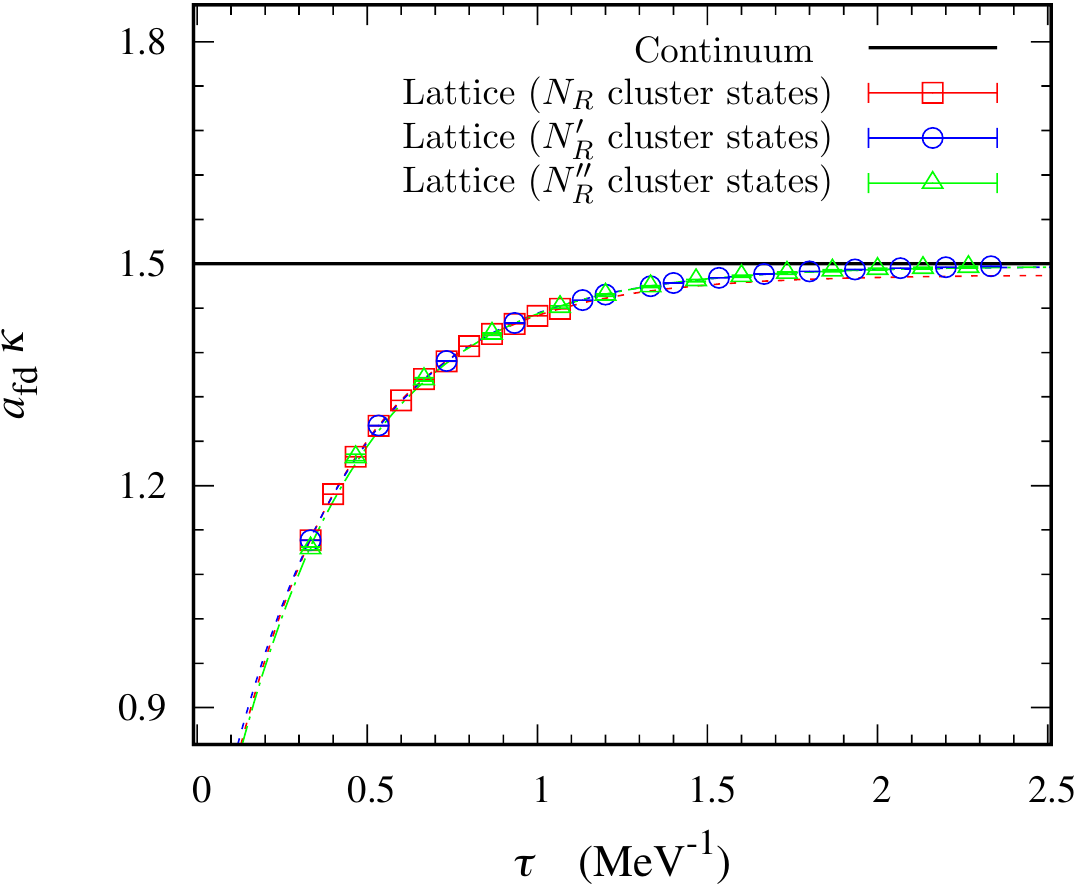}
\caption{Plot of the fermion-dimer scattering length as function of Euclidean time. The solid line denotes the result in the continuum, the red squares, blue circles, and green triangles are from the radial adiabatic Hamiltonians. $N_{R}$, $N'_{R}$, and  $N''_{R}$ denotes the number of initial cluster states given by Eq.~(\ref{eqn:initial-cluster-states-009}) for $m = 1$, $m = 2$, and $m = 3$, respectively. $\kappa$ is the dimer binding momentum $\kappa = \sqrt{2\mu \, B}$.}
\label{fig:afd_vs_Lt}
\end{figure}

We define the first radial adiabatic Hamiltonian in the subspace spanned by $N_{R}$ cluster states given by  Eq.~(\ref{eqn:initial-cluster-states-009}) for $m = 1$, the second radial adiabatic Hamiltonian in the subspace spanned by $N'_{R}$ cluster states given by  Eq.~(\ref{eqn:initial-cluster-states-009}) for $m = 2$, and the third radial adiabatic Hamiltonian in the subspace spanned by $N''_{R}$ cluster states given by  Eq.~(\ref{eqn:initial-cluster-states-009}) for $m = 3$.  We note that the eigenstates of each of these three radial adiabatic Hamiltonian form subspaces of the complete eigenstates of the microscopic Hamiltonian, and they are not orthogonal to the states lying outside the these subspaces. As discussed in Section~\ref{sec:two-particle-scattering}, by means of the Euclidean time projection method we can construct the adiabatic Hamiltonian as function of Euclidean time and project out the undesired states that are mixed with the low-energy states. Then we extract the two-cluster scattering information by applying the spherical wall method to the adiabatic Hamiltonian.

In fermion-dimer scattering, we compute the low-energy scattering phase shifts of the fermion-dimer system at various values of Euclidean time $\tau$. Since we are interested in scattering information at low energies, we find it convenient to present the fermion-dimer low-energy scattering parameters, \emph{i.e.} scattering length, instead of plotting the fermion-dimer scattering phase shifts at various values of $\tau$. Provided that the interaction between two clusters has a finite range, then at low energies the scattering phase shifts $\delta(p)$ can be parameterized by the effective range expansion,
\begin{align}
p \ \tan\delta(p) =
\frac{1}{a_{\rm fd}} + \frac{1}{2} \ r_{\rm fd} \ p^2 + \ldots \,,
\label{eqn:ERE-002}
\end{align}
where $a_{\rm fd}$ and $r_{\rm fd}$ are the fermion-dimer scattering length and effective range, respectively. Therefore, we construct the adiabatic Hamiltonian for various values of $\tau$ and use them to extract the two-cluster scattering phase shifts. Then we can compute the scattering length $a_{\rm fd}$ and the effective range $r_{\rm fd}$ at Euclidean time $\tau$ by fitting the truncated effective range expansion given in Eq.~(\ref{eqn:ERE-002}) to the lattice scattering data. In Fig.~\ref{fig:afd_vs_Lt} we plot ${a_{\rm fd}} \, \kappa$ as a function of Euclidean projection time $\tau$, where $\kappa$ is the dimer binding momentum $\kappa = \sqrt{2\mu \, B}$. These results show from the reduced adiabatic Hamiltonian, we can obtain the wave functions and use them in asymptotic limit to extract the low-energy scattering information. Furthermore, as seen in Fig.~\ref{fig:afd_vs_Lt}, the reduced adiabatic Hamiltonian calculations allows one to go larger Euclidean time since  it requires less numerical effort and also delays numerical stability problems produced by large ill-conditioned norm matrices.

\section{\textbf{Conclusion}}
\label{sec:conclusion}

The adiabatic projection method is a general framework for studying scattering and reactions on the lattice. Euclidean time projection plays a crucial role in the method since it works as a subspace projection and give information in the subspace which is spanned by the low-lying states of two-cluster system. 

In this paper we have presented the results from our investigations on extracting the low-energy eigenstates of two-body systems by diagonalizing the adiabatic Hamiltonian constructed in subspaces. We have examined the effects of reducing the size of the subspaces by studying two-particle systems in one and three spatial dimensions as well as fermion-dimer system in one spatial dimension. In two-particle systems, it is computationally practical to use every possible separation states, and the resulting adiabatic Hamiltonian matches with the microscopic Hamiltonian. Nevertheless, for simplicity and clarity we have studied two-particle systems deeply in one and three spatial dimensions, and we have constructed adiabatic Hamiltonians in subspaces of the total systems.  We have computed the two-particle scattering phase shifts in one dimension and two-particle scattering phase shifts for orbital angular momentum up to  $\ell = 4$ in three dimensions. We have found that the eigenstates of the adiabatic Hamiltonian are contaminated by the states lying outside the considered subspaces, and the Euclidean time projection method have released these contamination and given the exact description of the low-lying states for large Euclidean time.

Also we have extended our investigations from two-particle scattering to two-cluster scattering and considered a simplest non-trivial example of two-cluster system, the fermion-dimer system. We have constructed the radial adiabatic Hamiltonians using three different subspaces of the two-cluster state space. We have computed the fermion-dimer scattering phase shifts and extracted scattering parameters in the low-energy limit. Similarly, we have found that the eigenstates of the adiabatic Hamiltonian are mixed with the states which span outside the considered subspaces, and Euclidean time projection acts as a low-energy filter and cools the system to its low-lying states.

The results presented in this paper show that using a number of initial states which is less than the number of states of the complete system does not prevent us from extracting the low-lying eigenstates thanks to the Euclidean time projection method. Furthermore, due to the fact that using less number of initial cluster states requires smaller number of numerical operations, the numerical instability problem resulting from inverting large ill-conditional norm matrices is reduced significantly.

Throughout this paper we have employed exact methods for the adiabatic projection calculations, and the computational effort for exact methods scales quadratically with the number of initial cluster states $N_{R}$ but exponentially with nucleon number $A$. For instance, in order to compute the norm matrix  given in Eq.~(\ref{eqn:NormMatrix-001}), we need to calculate the dressed cluster states as shown in Eq.~(\ref{eqn:dressed-cluster-states-001}) by performing $L_{t}/2$ times matrix-matrix multiplications between the matrices $\exp(-Ha_{t}) \in \mathbb{C}^{L^{{\rm D}(A-1)}\times L^{{\rm D}(A-1)}}$ and $\ket{\bf R} \in \mathbb{C}^{L^{{\rm D}(A-1)}\times N_{R}}$, repeatedly. Here the spin and isospin degrees of freedom are neglected. The computation of the norm matrix in Eq.~(\ref{eqn:NormMatrix-001}) with exact methods requires $(2L^{{\rm D}(A-1)}-1)\, L^{{\rm D}(A-1)}\, N_{R}\, L_{t}/2 + (2L^{{\rm D}(A-1)}-1)\, N_{R}^{2}$ floating-point operations. Therefore, for a large number of nucleons the adiabatic projection calculations with the Euclidean time projection method can be employed  by means of the auxiliary-field projection Monte Carlo method~\cite{Elhatisari:2014lka,Elhatisari:2015iga,Elhatisari:2016owd}.  The calculation of the norm matrix using Monte Carlo methods requires 2$L^{{\rm D}}\, A\, (L_{t}+A) \, N_{R}^{2}$ floating-point operations, and the computational cost scales quadratically both with the number of initial cluster states $N_{R}$ and with nucleon number $A$. Therefore, being able to use a smaller $N_{R}$ would reduce the computational cost significantly, especially for large scale calculations. Therefore, the results and discussions given here provide a useful guide for improving the adiabatic projection method and for reducing the computational costs in lattice effective field theory calculations.

%%%%%%%%%%%%%%%%%%%%%%%%%%%%%%%%%%%%%%%%%%%%%%%%%%%%%%%%%%%%%%%%%%%
\section*{\textbf{Acknowledgment}} \label{sec:e}
The author is grateful to Dean~Lee and Ulf-G.~Mei{\ss}ner for stimulating discussions and for carefully reading the manuscript. This work is supported by the Scientific and Technological Research Council of Turkey (TUBITAK) project no. 116F400.


\begin{thebibliography}{1}


%%%%%%%%%%%%%%%%%%%%%%%%%%%%%%%%%%%%%%%%%%%%%%%%%%%%%%%%%%%%%%%%%%%%%%%%

%\cite{Weaver:1993zz}
\bibitem{Weaver:1993zz} 
  T.~A.~Weaver and S.~E.~Woosley,
  %``Nucleosynthesis in massive stars and the C-12(alpha,gamma)O-16 reaction rate,''
  Phys.\ Rept.\  {\bf 227}, 65 (1993).
  doi:10.1016/0370-1573(93)90058-L
  %%CITATION = doi:10.1016/0370-1573(93)90058-L;%%
  %186 citations counted in INSPIRE as of 07 May 2019

%\cite{Adelberger:2010qa}
\bibitem{Adelberger:2010qa} 
  E.~G.~Adelberger {\it et al.},
  %``Solar fusion cross sections II: the pp chain and CNO cycles,''
  Rev.\ Mod.\ Phys.\  {\bf 83}, 195 (2011)
  doi:10.1103/RevModPhys.83.195
  [arXiv:1004.2318 [nucl-ex]].
  %%CITATION = doi:10.1103/RevModPhys.83.195;%%
  %327 citations counted in INSPIRE as of 08 May 2019

%\cite{Leva:2016wmg}
\bibitem{Leva:2016wmg} 
  A.~D.~Leva, L.~Gialanella and F.~Strieder,
  %``Experimental status of 7Be production and destruction at astrophysical relevant energies,''
  J.\ Phys.\ Conf.\ Ser.\  {\bf 665}, no. 1, 012002 (2016).
  doi:10.1088/1742-6596/665/1/012002
  %%CITATION = doi:10.1088/1742-6596/665/1/012002;%%
  %1 citations counted in INSPIRE as of 08 May 2019


%\cite{Nollett:2006su}
\bibitem{Nollett:2006su} 
  K.~M.~Nollett, S.~C.~Pieper, R.~B.~Wiringa, J.~Carlson and G.~M.~Hale,
  %``Quantum Monte Carlo calculations of neutron-alpha scattering,''
  Phys.\ Rev.\ Lett.\  {\bf 99}, 022502 (2007)
  doi:10.1103/PhysRevLett.99.022502
  [nucl-th/0612035].
  %%CITATION = doi:10.1103/PhysRevLett.99.022502;%%
  %142 citations counted in INSPIRE as of 07 May 2019

%\cite{Navratil:2011zs}
\bibitem{Navratil:2011zs} 
  P.~Navratil and S.~Quaglioni,
  %``Ab initio many-body calculations of the 3H(d,n)4He and 3He(d,p)4He fusion,''
  Phys.\ Rev.\ Lett.\  {\bf 108}, 042503 (2012)
  doi:10.1103/PhysRevLett.108.042503
  [arXiv:1110.0460 [nucl-th]].
  %%CITATION = doi:10.1103/PhysRevLett.108.042503;%%
  %76 citations counted in INSPIRE as of 07 May 2019

%\cite{Hagen:2012rq}
\bibitem{Hagen:2012rq} 
  G.~Hagen and N.~Michel,
  %``Elastic proton scattering of medium mass nuclei from coupled-cluster theory,''
  Phys.\ Rev.\ C {\bf 86}, 021602 (2012)
  doi:10.1103/PhysRevC.86.021602
  [arXiv:1206.2336 [nucl-th]].
  %%CITATION = doi:10.1103/PhysRevC.86.021602;%%
  %48 citations counted in INSPIRE as of 07 May 2019

%\cite{Quaglioni:2013kma}
\bibitem{Quaglioni:2013kma} 
  S.~Quaglioni, C.~Romero-Redondo and P.~Navratil,
  %``Three-cluster dynamics within an ab initio framework,''
  Phys.\ Rev.\ C {\bf 88}, 034320 (2013)
  Erratum: [Phys.\ Rev.\ C {\bf 94}, no. 1, 019902 (2016)]
  doi:10.1103/PhysRevC.94.019902, 10.1103/PhysRevC.88.034320
  [arXiv:1307.8160 [nucl-th]].
  %%CITATION = doi:10.1103/PhysRevC.94.019902, 10.1103/PhysRevC.88.034320;%%
  %21 citations counted in INSPIRE as of 07 May 2019

%\cite{Elhatisari:2015iga}
\bibitem{Elhatisari:2015iga} 
  S.~Elhatisari, D.~Lee, G.~Rupak, E.~Epelbaum, H.~Krebs, T.~A.~L\"{a}hde, T.~Luu and Ulf-G.~Mei{\ss}ner,
  %``Ab initio alpha-alpha scattering,''
  Nature {\bf 528}, 111 (2015)
  doi:10.1038/nature16067
  [arXiv:1506.03513 [nucl-th]].
  %%CITATION = doi:10.1038/nature16067;%%
  %57 citations counted in INSPIRE as of 07 May 2019

%\cite{Navratil:2016ycn}
\bibitem{Navratil:2016ycn} 
  P.~Navratil, S.~Quaglioni, G.~Hupin, C.~Romero-Redondo and A.~Calci,
  %``Unified ab initio approaches to nuclear structure and reactions,''
  Phys.\ Scripta {\bf 91}, no. 5, 053002 (2016)
  doi:10.1088/0031-8949/91/5/053002
  [arXiv:1601.03765 [nucl-th]].
  %%CITATION = doi:10.1088/0031-8949/91/5/053002;%%
  %57 citations counted in INSPIRE as of 07 May 2019


%\cite{Dohet-Eraly:2015ooa}
\bibitem{Dohet-Eraly:2015ooa} 
  J.~Dohet-Eraly, P.~Navratil, S.~Quaglioni, W.~Horiuchi, G.~Hupin and F.~Raimondi,
  %``$^3$He($\alpha,\gamma$)$^7$Be and $^3$H($\alpha,\gamma$)$^7$Li astrophysical S factors from the no-core shell model with continuum,''
  Phys.\ Lett.\ B {\bf 757}, 430 (2016)
  doi:10.1016/j.physletb.2016.04.021
  [arXiv:1510.07717 [nucl-th]].
  %%CITATION = doi:10.1016/j.physletb.2016.04.021;%%
  %18 citations counted in INSPIRE as of 07 May 2019



%\cite{Rupak:2013aue}
\bibitem{Rupak:2013aue} 
  G.~Rupak and D.~Lee,
  %``Radiative capture reactions in lattice effective field theory,''
  Phys.\ Rev.\ Lett.\  {\bf 111}, no. 3, 032502 (2013)
  doi:10.1103/PhysRevLett.111.032502
  [arXiv:1302.4158 [nucl-th]].
  %%CITATION = doi:10.1103/PhysRevLett.111.032502;%%
  %31 citations counted in INSPIRE as of 08 May 2019

%\cite{Pine:2013zja}
\bibitem{Pine:2013zja} 
  M.~Pine, D.~Lee and G.~Rupak,
  %``Adiabatic projection method for scattering and reactions on the lattice,''
  Eur.\ Phys.\ J.\ A {\bf 49}, 151 (2013)
  doi:10.1140/epja/i2013-13151-3
  [arXiv:1309.2616 [nucl-th]].
  %%CITATION = doi:10.1140/epja/i2013-13151-3;%%
  %25 citations counted in INSPIRE as of 08 May 2019

%\cite{Elhatisari:2014lka}
\bibitem{Elhatisari:2014lka} 
  S.~Elhatisari and D.~Lee,
  %``Fermion-dimer scattering using an impurity lattice Monte Carlo approach and the adiabatic projection method,''
  Phys.\ Rev.\ C {\bf 90}, no. 6, 064001 (2014)
  doi:10.1103/PhysRevC.90.064001
  [arXiv:1407.2784 [nucl-th]].
  %%CITATION = doi:10.1103/PhysRevC.90.064001;%%
  %10 citations counted in INSPIRE as of 08 May 2019

%\cite{Rupak:2014xza}
\bibitem{Rupak:2014xza} 
  G.~Rupak and P.~Ravi,
  %``Proton–proton fusion in lattice effective field theory,''
  Phys.\ Lett.\ B {\bf 741}, 301 (2015)
  doi:10.1016/j.physletb.2014.12.055
  [arXiv:1411.2436 [nucl-th]].
  %%CITATION = doi:10.1016/j.physletb.2014.12.055;%%
  %5 citations counted in INSPIRE as of 08 May 2019


%\cite{Rokash:2015hra}
\bibitem{Rokash:2015hra} 
  A.~Rokash, M.~Pine, S.~Elhatisari, D.~Lee, E.~Epelbaum and H.~Krebs,
  %``Scattering cluster wave functions on the lattice using the adiabatic projection method,''
  Phys.\ Rev.\ C {\bf 92}, no. 5, 054612 (2015)
  doi:10.1103/PhysRevC.92.054612
  [arXiv:1505.02967 [nucl-th]].
  %%CITATION = doi:10.1103/PhysRevC.92.054612;%%
  %7 citations counted in INSPIRE as of 08 May 2019

%\cite{Elhatisari:2016hby}
\bibitem{Elhatisari:2016hby} 
  S.~Elhatisari, D.~Lee, Ulf-G.~Mei{\ss}ner and G.~Rupak,
  %``Nucleon-deuteron scattering using the adiabatic projection method,''
  Eur.\ Phys.\ J.\ A {\bf 52}, no. 6, 174 (2016)
  doi:10.1140/epja/i2016-16174-2
  [arXiv:1603.02333 [nucl-th]].
  %%CITATION = doi:10.1140/epja/i2016-16174-2;%%
  %9 citations counted in INSPIRE as of 08 May 2019


%\cite{Elhatisari:2016owd}
\bibitem{Elhatisari:2016owd} 
  S.~Elhatisari {\it et al.},
  %``Nuclear binding near a quantum phase transition,''
  Phys.\ Rev.\ Lett.\  {\bf 117}, no. 13, 132501 (2016)
  doi:10.1103/PhysRevLett.117.132501
  [arXiv:1602.04539 [nucl-th]].
  %%CITATION = doi:10.1103/PhysRevLett.117.132501;%%
  %27 citations counted in INSPIRE as of 08 May 2019
  
  
  \bibitem{lahde2019nuclear} 
  T.~A.~L\"ahde and U.-G.~Mei{\ss}ner,
  %``Nuclear Lattice Effective Field Theory : An introduction,''
  Lect.\ Notes Phys.\  {\bf 957}, 1 (2019).
  doi:10.1007/978-3-030-14189-9
  %%CITATION = doi:10.1007/978-3-030-14189-9;%%

%\cite{Lu:2014xfa}
\bibitem{Lu:2014xfa} 
  B.~N.~Lu, T.~A.~L\"{a}hde, D.~Lee and Ulf-G.~Mei{\ss}ner,
  %``Breaking and restoration of rotational symmetry on the lattice for bound state multiplets,''
  Phys.\ Rev.\ D {\bf 90}, no. 3, 034507 (2014)
  doi:10.1103/PhysRevD.90.034507
  [arXiv:1403.8056 [nucl-th]].
  %%CITATION = doi:10.1103/PhysRevD.90.034507;%%
  %14 citations counted in INSPIRE as of 09 May 2019

%\cite{Stellin:2018fkj}
\bibitem{Stellin:2018fkj} 
  G.~Stellin, S.~Elhatisari and Ulf-G.~Mei{\ss}ner,
  %``Breaking and restoration of rotational symmetry in the low-energy spectrum of light alpha-conjugate nuclei on the lattice I: $^{8}\mathrm{Be}$ and $^{12}\mathrm{C}$,''
  Eur.\ Phys.\ J.\ A {\bf 54}, no. 12, 232 (2018)
  doi:10.1140/epja/i2018-12671-6
  [arXiv:1809.06109 [nucl-th]].
  %%CITATION = doi:10.1140/epja/i2018-12671-6;%%


%\cite{Lu:2015riz}
\bibitem{Lu:2015riz} 
  B.~N.~Lu, T.~A.~L\"{a}hde, D.~Lee and Ulf-G.~Mei{\ss}ner,
  %``Precise determination of lattice phase shifts and mixing angles,''
  Phys.\ Lett.\ B {\bf 760}, 309 (2016)
  doi:10.1016/j.physletb.2016.06.081
  [arXiv:1506.05652 [nucl-th]].
  %%CITATION = doi:10.1016/j.physletb.2016.06.081;%%
  %17 citations counted in INSPIRE as of 10 May 2019
  
%\cite{Borasoy:2007vy}
\bibitem{Borasoy:2007vy} 
  B.~Borasoy, E.~Epelbaum, H.~Krebs, D.~Lee and Ulf-G.~Meissner,
  %``Two-particle scattering on the lattice: Phase shifts, spin-orbit coupling, and mixing angles,''
  Eur.\ Phys.\ J.\ A {\bf 34}, 185 (2007)
  doi:10.1140/epja/i2007-10500-9
  [arXiv:0708.1780 [nucl-th]].
  %%CITATION = doi:10.1140/epja/i2007-10500-9;%%
  %43 citations counted in INSPIRE as of 18 May 2019
  
  


\end{thebibliography}
\end{document}